# Temporal User Profiling with LLMs: Balancing Short-Term and Long-Term Preferences for Recommendations

MILAD SABOURI, DePaul University, USA

MASOUD MANSOURY, Delft University of Technology, Netherlands

KUN LIN, DePaul University, USA

BAMSHAD MOBASHER, DePaul University, USA

Accurately modeling user preferences is crucial for improving the performance of content-based recommender systems. Existing approaches often rely on simplistic user profiling methods, such as averaging or concatenating item embeddings, which fail to capture the nuanced nature of user preference dynamics, particularly the interactions between long-term and short-term preferences. In this work, we propose LLM-driven Temporal User Profiling (LLM-TUP), a novel method for user profiling that explicitly models short-term and long-term preferences by leveraging interaction timestamps and generating natural language representations of user histories using a large language model (LLM). These representations are encoded into high-dimensional embeddings using a pre-trained BERT model, and an attention mechanism is applied to dynamically fuse the short-term and long-term embeddings into a comprehensive user profile. Experimental results on real-world datasets demonstrate that LLM-TUP achieves substantial improvements over several baselines, underscoring the effectiveness of our temporally aware user-profiling approach and the use of semantically rich user profiles, generated by LLMs, for personalized content-based recommendation.

## 1 INTRODUCTION

In content-based recommender systems, a common approach for creating user embeddings involves an aggregation-based approach where item embeddings are first generated from textual information such as titles or descriptions. Subsequently, user embeddings are computed as the average of the embeddings of all items interacted with by a user [2, 17, 28]. While straightforward and computationally efficient, this approach overlooks critical aspects of user behavior, resulting in suboptimal user profiles. For simplicity and consistency, we refer to this method of creating user profiling as the "Centric" approach throughout this document.

This centric approach suffers a number of limitations. First, it fails to distinguish users' short-term and long-term preferences. User interactions are often influenced by transient contexts, such as seasonal trends [18] or recent events [22], which are diluted in the centric approach. For instance, a user's interest in holiday-themed content during a specific period is averaged together with their broader long-term preferences, failing to reflect the temporary change in behavior.

Second, the centric approach may confuse the semantic similarity between items with which the user has interacted with. Users who engage with items in different contexts could end up with preferences that seem overly uniform [37]. For example, a user who likes both action movies and romantic comedies might be represented by a single embedding that fails to capture the variety and balance of their interests.

Third, user preferences often evolve over time and are influenced by external factors such as trends and temporal contexts. In domains like movies or music [12, 25], user behavior patterns can exhibit significant variability. For example, holiday-specific content may dominate user preferences during festive seasons, while differences in weekday versus

Authors' Contact Information: Milad Sabouri, DePaul University, Chicago, USA, msabouri@depaul.edu; Masoud Mansoury, Delft University of Technology, Delft, Netherlands, m.mansoury@tudelft.nl; Kun Lin, DePaul University, Chicago, USA, klin13@depaul.edu; Bamshad Mobasher, DePaul University, Chicago, USA, mobasher@cs.depaul.edu.

weekend behaviors can affect the types of items users engage with. Ignoring these temporal dynamics restricts the recommender system's ability to adapt to shifts in user interests.

Fourth, while the centric approach provides a baseline for generating user embeddings, its structural simplicity makes it ill-suited for modeling users with sparse interaction histories or diverse interests. Users with limited data points are particularly at risk of being represented by embeddings that fail to generalize effectively. Similarly, users whose interactions span multiple, non-cohesive categories may be misrepresented by a single averaged embedding. These challenges highlight the need for a more dynamic and context-aware methodology.

To address these limitations in creating user embeddings for content-based recommender systems, we propose LLM-driven Temporal User Profiling (LLM-TUP), a novel method that enhances user profiling by incorporating temporal dynamics and semantic richness into user representations. The name LLM-TUP reflects the dual focus of the method: leveraging temporal information from user interaction histories and employing Large Language Models (LLMs) to generate context-aware user profiles. In this work, our research focuses exclusively on improving **user profiling**, demonstrating that a better representation of user preferences leads to improved recommendation performance. Through extensive experiments and ablation studies, we show that LLM-TUP significantly outperforms the centric approach, a widely used method for content-based user modeling, in effectively capturing both short-term and long-term user preferences.

This study aims to answer three key research questions:

- **RQ1**: Does our proposed user modeling method, LLM-TUP, result in more accurate ranking performance for recommendations compared to other baselines?
- **RQ2**: How effective is LLM in generating semantically rich and temporally aware profiles for short-term and long-term preferences, compared to traditional content-based approaches?
- **RQ3**: How do different components of LLM-TUP contribute to improving recommendation performance?

To address these questions, LLM-TUP introduces the following key innovations:

- Temporal Representation via LLM: Using an LLM to transform interaction histories and timestamps into two natural language representations that encapsulate short-term and long-term preferences.
- Dynamic Fusion with Attention Mechanism: Leveraging an attention mechanism to dynamically determine the relative importance of short-term and long-term preferences for each user, ensuring personalized embeddings.
- End-to-End Integration: Combining user embeddings with item embeddings for recommendation prediction, enhancing adaptability to various user behaviors.

By explicitly modeling temporal dynamics and leveraging LLM-based profiling, LLM-TUP establishes a more effective approach to user modeling.

## 2 RELATED WORKS

### 2.1 User Modeling in Recommender Systems

User modeling is a fundamental component of recommender systems [3, 7, 9, 15], aiming to represent user preferences based on their interactions with items. Traditional collaborative filtering (CF) approaches, such as matrix factorization [16], represent users and items in a shared latent space using interaction data. However, CF methods often struggle in sparse data scenarios or when dealing with new users or items [14, 24]. To address these limitations, content-based recommendation methods [20] have gained prominence by leveraging item metadata, such as textual descriptions or user reviews, to generate semantically meaningful representations.

Sequential recommendation models, such as BERT4Rec [33], SASRec [13], and GRU4Rec [34] explicitly model the order of interactions to predict the next item in a sequence. While effective for session-based and sequential recommendations, these approaches rely on collaborative filtering signals and do not leverage textual item metadata for user profiling. In contrast, our work focuses on content-based user profiling, where we generate semantically rich user representations using LLMs while incorporating temporal information into user profiles.

## 2.2 Content-Based Recommendations

Content-based recommendation systems rely on item metadata (e.g., descriptions, titles, or reviews) to model user preferences. Early methods extracted handcrafted features from textual content [2, 17], but deep learning has since enabled automated feature extraction via Deep Content-Based Recommendation (DCBR) [21, 36]. Pre-trained language models like BERT [6] have further enhanced content-based recommendations by generating contextualized item embeddings, improving personalization [27]. However, most existing content-based approaches fail to incorporate temporal dynamics, treating all user interactions equally, regardless of when they occurred [4]. Our work addresses this limitation by explicitly modeling short-term and long-term user preferences, generating temporal-aware user profiles with LLMs.

## 2.3 Large Language Models in Recommender Systems

The rise of Large Language Models (LLMs) has opened new avenues for personalized recommendations [32, 35, 38, 41]. LLMs have been leveraged to generate user and item embeddings by processing textual content, such as user reviews, item descriptions, or interaction histories. [5] refines sequential user interaction sequences by aligning user interaction sequences with LLM-generated embeddings. [31] explores lightweight yet effective LLM-based recommendation techniques, focusing on integrating LLMs into collaborative filtering models. [19] develops an LLM-driven fashion recommendation framework, demonstrating LLM effectiveness in session-based recommendations. LLMs have also been employed in zero-shot recommendation, leveraging pre-trained knowledge to generalize across domains [11].

While these works highlight the effectiveness of LLMs in recommendation, they primarily focus on sequential modeling and collaborative filtering signals, rather than content-based user profiling. Our method is fundamentally different as it utilizes LLMs to process user interaction histories, generating natural language representations of short-term and long-term preferences and dynamically fusing them with attention mechanisms for temporally aware content-based recommendations.

## 2.4 Temporal Dynamics in Recommendations

Temporal dynamics play a critical role in understanding user preferences. Sequential recommendation models like GRU4Rec [34], SASRec [13], and Time-LSTM [42] have demonstrated the importance of modeling temporal dependencies in user interactions. However, these methods rely primarily on collaborative filtering signals and are designed to model sequences of interactions without leveraging item metadata. More recent works have attempted to refine how long-term and short-term user preferences are integrated. [40] introduces a time-aware and content-aware recurrent model that adaptively fuses short-term and long-term user behaviors. Similarly, [23] proposes a self-attention and BiGRU hybrid recommendation model to extract short-term and long-term preferences before fusing them into a unified user representation.

In summary, our work builds on advances in content-based recommendation, LLM-based representation learning, and temporal modeling. While existing CF-based temporal models like SASRec and GRU4Rec focus on interaction sequences, our method uniquely integrates temporal dynamics with content-based signals using LLMs and BERT. This novel approach addresses key limitations of prior works, providing semantically rich and temporally aware user profiles that enhance recommendation quality in content-based systems.

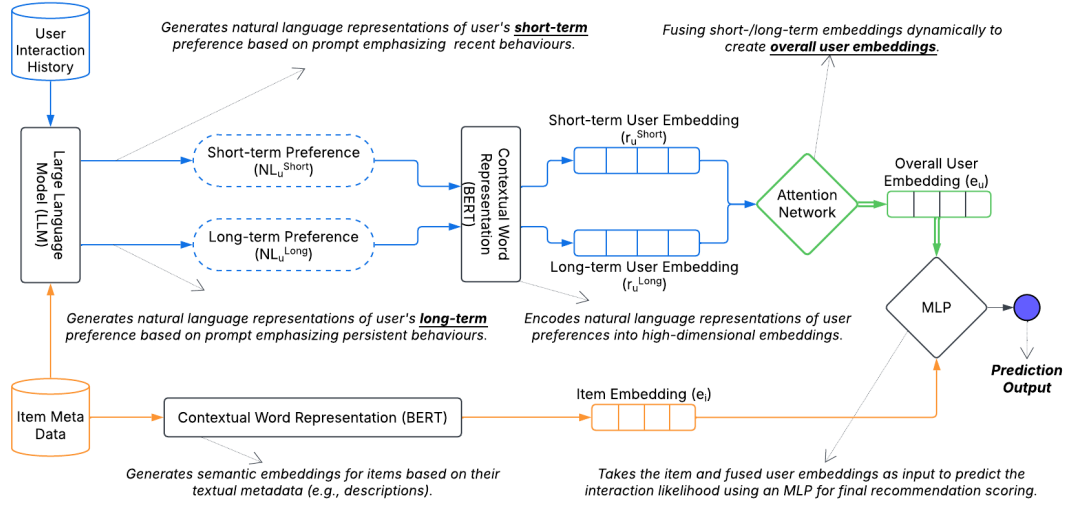

Fig. 1. Proposed Architecture for LLM-Driven Temporal User Profiling

## 3  The Proposed LLM-TUP Framework

Let $\mathcal{U} = \{u_1, u_2, \ldots, u_{|\mathcal{U}|}\}$ denote the set of users, and $\mathcal{I} = \{i_1, i_2, \ldots, i_{|\mathcal{I}|}\}$ denote the set of items. Each user $u \in \mathcal{U}$ interacts with a subset of items $\mathcal{I}_u \subseteq \mathcal{I}$, where each interaction is associated with a timestamp $t_{u,i}$. Let $\mathcal{H}_u = \{(i, t_{u,i}) : i \in \mathcal{I}_u\}$ represent the interaction history of user $u$, sorted in ascending order of timestamps.

The objective is to learn a *user embedding* $\mathbf{e}_u \in \mathbb{R}^d$ and an *item embedding* $\mathbf{e}_i \in \mathbb{R}^d$ for each user $u$ and item $i$, such that the likelihood of user $u$ interacting with item $i$ can be predicted accurately. Specifically, we aim to estimate a function $f : \mathbb{R}^d \times \mathbb{R}^d \rightarrow [0, 1]$ that predicts the interaction probability $\hat{y}_{u,i}$ as

$$\hat{y}_{u,i} = f(\mathbf{e}_u, \mathbf{e}_i), \tag{1}$$

where $\hat{y}_{u,i}$ denotes the predicted likelihood of user $u$ interacting with item $i$, and $f$ is implemented as a multi-layer perceptron (MLP).

The proposed architecture for LLM-driven temporal user profiling, illustrated in Figure 1, leverages temporal information and semantic representations to generate personalized user embeddings. The model consists of three key components: (i) Temporal User Profile Creation, (ii) Embedding Representation, and (iii) Recommendation Generation.

### 3.1 LLM-Driven Temporal User Profile Creation

To construct a temporal-aware user profile, we generate two distinct **natural language representations** of user preferences: *short-term preferences* and *long-term preferences*. Let $\mathcal{H}_u = \{(i, t_{u,i}) : i \in \mathcal{I}_u\}$ denote the interaction history of user $u$, where $t_{u,i}$ represents the timestamp of interaction with item $i$. The interaction history is sorted in ascending order of timestamps to maintain chronological order.

We utilize a Large Language Model (LLM) to process the entire user interaction history twice, using distinct prompts to generate separate representations for short-term and long-term preferences:

*Pass 1: Short-Term Profile Generation* – The complete interaction history $\mathcal{H}_u$ is provided as input to the LLM along with a task-specific prompt that instructs the model to extract the user's *short-term interests*, placing greater emphasis on the most recent interactions while still considering temporal context. The LLM generates a natural language representation of short-term preferences, denoted as:

$$\mathrm{NL}_u^{\mathrm{short}} = \mathrm{LLM}(\mathcal{H}_u, \mathrm{Prompt}^{\mathrm{short}}) \qquad (2)$$

*Pass 2: Long-Term Profile Generation* – The same interaction history $\mathcal{H}_u$ is passed to the LLM again, but this time with a different prompt that instructs the model to generate a *long-term preference profile* by considering the user's entire history, capturing overarching interests and persistent behavioral patterns. The resulting representation is:

$$\mathrm{NL}_u^{\mathrm{long}} = \mathrm{LLM}(\mathcal{H}_u, \mathrm{Prompt}^{\mathrm{long}}) \qquad (3)$$

By explicitly utilizing distinct prompts while keeping the full interaction history intact, this approach enables the LLM to contextually differentiate between recent and persistent user preferences, addressing the limitations of the traditional centric-based approach that fails to incorporate temporal user dynamics.

### 3.2 Embedding Representation

The generated natural language representations are encoded into high-dimensional embeddings using a pre-trained BERT model [6], resulting in short-term user embedding and long-term user embedding, as follows:

$$\mathbf{r}_u^{\mathrm{short}} = \mathrm{BERT}(\mathrm{NL}_u^{\mathrm{short}}), \qquad (4)$$

$$\mathbf{r}_u^{\mathrm{long}} = \mathrm{BERT}(\mathrm{NL}_u^{\mathrm{long}}), \qquad (5)$$

where $\mathbf{r}_u^{\mathrm{short}} \in \mathbb{R}^d$ and $\mathbf{r}_u^{\mathrm{long}} \in \mathbb{R}^d$ are $d$-dimensional embeddings representing the short-term and long-term preferences of user $u$, respectively.

Equations 4 and 5 describe the encoding process where BERT converts the natural language profiles into dense vector representations. These embeddings capture both the semantic richness of user preferences and contextual nuances from interaction histories, serving as the basis for constructing the final user profile. Similarly, item descriptions are processed by BERT to generate item embeddings, ensuring that user and item representations lie in the same embedding space.

To construct the overall user embedding, an attention mechanism [39] is applied to dynamically fuse the short-term and long-term user embeddings. The attention mechanism computes personalized importance weights for short-term and long-term preferences, enabling the model to adaptively emphasize recent interactions for users with rapidly changing interests or long-term behavior for users with stable preferences. This dynamic fusion mechanism is a key differentiator from the centric approach, where user profiles are created by simply averaging item embeddings, thereby neglecting temporal nuances.

Let $\alpha_u^{\text{short}}$ and $\alpha_u^{\text{long}}$ denote the attention weights assigned to the short-term and long-term preference embeddings, respectively. The attention weights are computed as:

$$\alpha_u^{\text{short}} = \frac{\exp(\mathbf{W}_a \mathbf{r}_u^{\text{short}})}{\exp(\mathbf{W}_a \mathbf{r}_u^{\text{short}}) + \exp(\mathbf{W}_a \mathbf{r}_u^{\text{long}})}, \tag{6}$$

$$\alpha_u^{\text{long}} = 1 - \alpha_u^{\text{short}}, \tag{7}$$

where $\mathbf{W}_a \in \mathbb{R}^{1 \times d}$ is a learnable parameter vector, and $\exp(\cdot)$ denotes the exponential function. Equation 6 computes the attention weight for short-term preferences, while Equation 7 ensures that the sum of the attention weights is equal to 1.

The overall user embedding $\mathbf{e}_u$ is then obtained as a weighted sum of the short-term and long-term embeddings:

$$\mathbf{e}_u = \alpha_u^{\text{short}} \cdot \mathbf{r}_u^{\text{short}} + \alpha_u^{\text{long}} \cdot \mathbf{r}_u^{\text{long}}, \tag{8}$$

where $\mathbf{e}_u \in \mathbb{R}^d$ represents the final user embedding. As shown in Equation 8, the attention mechanism enables the model to dynamically balance the influence of short-term and long-term preferences based on user-specific interaction patterns.

By learning the attention weights $\alpha_u^{\text{short}}$ and $\alpha_u^{\text{long}}$, the model can adaptively prioritize recent interactions for users with rapidly changing interests or emphasize long-term preferences for users with stable behavior.

### 3.3 Recommendation Generation

The overall user embedding $\mathbf{e}_u$, obtained from the attention mechanism, is concatenated with the corresponding item embedding $\mathbf{e}_i$ and passed through a multi-layer perceptron (MLP) [29] to predict the interaction probability of user $u$:

$$\hat{y}_{u,i} = \text{MLP}([\mathbf{e}_u; \mathbf{e}_i]), \tag{9}$$

where $[\mathbf{e}_u; \mathbf{e}_i] \in \mathbb{R}^{2d}$ denotes the concatenation of the user embedding $\mathbf{e}_u$ and the item embedding $\mathbf{e}_i$, and $\hat{y}_{u,i} \in [0, 1]$ represents the predicted probability that user $u$ will interact with item $i$.

As shown in Equation 9, the MLP serves as a scoring function that learns non-linear interactions between user and item embeddings. The final output is a probability value indicating the likelihood of the user interacting with the given item. The MLP consists of multiple fully connected layers with ReLU activation functions, followed by a sigmoid output layer to ensure that the output lies in the range $[0, 1]$.

The model is trained using a binary cross-entropy loss, where the target label $y_{u,i} \in \{0, 1\}$ indicates whether user $u$ interacted with item $i$:

$$\mathcal{L} = -\frac{1}{|\mathcal{D}|} \sum_{(u,i,y_{u,i}) \in \mathcal{D}} \left( y_{u,i} \log \hat{y}_{u,i} + (1 - y_{u,i}) \log(1 - \hat{y}_{u,i}) \right), \tag{10}$$

where $\mathcal{D}$ represents the set of all user-item interaction pairs in the training dataset. As shown in Equation 10, the binary cross-entropy loss penalizes the model based on the difference between the predicted interaction probability $\hat{y}_{u,i}$ and the true label $y_{u,i}$. The objective is to minimize this loss function, thereby ensuring that the model accurately predicts interactions.

Table 1. Datasets Statistics

| Dataset | # Unique Users | # Unique Items | # Interactions | Avg. User Profile Size |
|---------|----------------|----------------|----------------|------------------------|
| Movies&TV | 10,000 | 14,420 | 202,583 | 10.28 |
| Games | 10,371 | 3,790 | 83,842 | 4.55 |

Compared to the baseline (referred to as the centric approach), which generates user profiles by averaging item embeddings without distinguishing between short-term and long-term preferences, the proposed architecture offers two key advantages:

- *Temporal Awareness:* By explicitly modeling short-term and long-term preferences and dynamically fusing them using an attention mechanism, the model can better capture the evolving nature of user behavior.
- *Semantic Richness:* The use of LLM-generated natural language profiles followed by BERT encoding, ensures that user embeddings are semantically rich and contextually meaningful. This leads to better alignment between user and item representations, and, consequently, more accurate recommendations.

The combination of temporal modeling, attention-based fusion, and semantic representations enables the proposed model to outperform the centric approach in terms of recommendation accuracy, as evidenced by significant improvements in ranking metrics such as Recall@K and NDCG@K.

## 4   Experiments

In this research, **our primary focus** is to propose a novel method for creating **user profiles**, specifically for content-based recommendation systems. The goal of these experiments is to evaluate the effectiveness of the proposed method, LLM-TUP, in incorporating temporal dynamics, improving recommendation quality, and generating richer user profiles compared to baseline models.

### 4.1   Datasets

We evaluated our proposed method using two datasets from Amazon Product Review [10], focusing on the Movies&TV and Video Games categories. These datasets contain interaction data, including timestamps for each user-item interaction, enabling the incorporation of temporal dynamics. Additionally, metadata for each item, such as titles and descriptions, was utilized to generate item embeddings. Specifically, we applied BERT to the textual descriptions of items to create feature-rich embeddings that serve as input to both the baselines and the proposed method. Table 1 presents the dataset statistics.

### 4.2   Baselines

Given our focus on user profiling within content-based recommendation systems, it is more relevant to compare our proposed method against the user profiling approaches typically employed in such systems. To assess the effectiveness of our method, we benchmark it against several baselines that encompass a range of user modeling and recommendation strategies. These baselines were selected to illuminate various aspects of user profiling, from simple heuristic-based techniques to more sophisticated modeling approaches. By including these diverse baselines, we can evaluate the impact of incorporating temporal information, large language models (LLMs), and attention-based fusion in creating personalized user profiles.

*4.2.1  Centric Approach (Centric).*  The centric approach serves as the primary baseline for comparison. In this method, user profiles are constructed by averaging the BERT embeddings of all items the user has interacted with, irrespective of the timestamps of interactions. To ensure a fair comparison, other components of the pipeline, including the item embeddings and MLP, were kept identical to the proposed method. This approach represents the common practice [8] of aggregating item embeddings to approximate user preferences, making it a natural and relevant baseline for evaluating the benefits of introducing temporal information and dynamic fusion. By comparing against this method, we can quantify the improvements achieved by explicitly modeling short-term and long-term preferences.

*4.2.2  Temporal Fusion without LLM-Based Representations (Temp-Fusion).*  This baseline represents a traditional content-based recommendation approach that incorporates temporal dynamics without using LLM-generated representations. It constructs user profiles by dividing the user's interaction history into short-term and long-term segments based on timestamps. The cutoff value that determines how many of the most recent interactions are considered short-term interactions is treated as a hyperparameter and varies depending on dataset characteristics. The results presented in this study use a cutoff of 3 interactions for the Movies dataset and a cutoff of 1 interaction for the Games dataset, based on empirical tuning. Each segment is represented using BERT embeddings derived from item metadata (e.g., descriptions, titles), following conventional content-based profiling techniques. To capture both recent interests and persistent preferences, the short-term and long-term profiles are fused using an attention network, similar to our proposed method. The resulting fused user profile is then passed to an MLP, along with item embeddings, to predict interaction likelihood. Unlike our proposed method, Temp-Fusion does not utilize LLM-generated natural language representations. Instead, it relies purely on the raw embeddings of interacted items, making it a strong content-based baseline. This comparison isolates the contribution of LLM-based semantic profiling, demonstrating its impact on improving recommendation performance.

*4.2.3  Popularity-Based Recommendation (Popularity).*  The popularity-based recommendation baseline ranks items based on their global popularity, measured by the frequency of interactions across all users. This non-personalized baseline serves as a reference point to illustrate the value of personalization in recommendation. While simplistic, it provides a lower bound on performance and highlights the need for sophisticated user profiling methods.

*4.2.4  Matrix Factorization (MF).*  Matrix Factorization is included as a baseline to compare against a widely-used collaborative filtering (CF) method. MF generates latent user and item factors by factorizing the interaction matrix, relying solely on collaborative signals. Although our method focuses on content-based user profiling rather than CF, this baseline is relevant for understanding the differences between CF-based and content-based approaches. By comparing the performance of MF with our method, we can demonstrate how leveraging item metadata and temporal information enables effective personalization in contexts where collaborative signals may be sparse or unavailable.

These baselines were carefully selected to evaluate the contribution of key components of our proposed method, including temporal modeling, semantic representation using LLMs, and attention-based fusion. By comparing against these baselines, we demonstrate that our method's ability to generate temporally aware, semantically rich, and personalized user profiles significantly enhances recommendation performance.

Table 2. Performance Comparison our LLM-TUP method and baselines on Movies&TV and Video Games datasets for $K = 10$ and $K = 20$.

| Method | Movies&TV | | | | Video Games | | | |
|---|---|---|---|---|---|---|---|---|
| | Recall@10 | NDCG@10 | Recall@20 | NDCG@20 | Recall@10 | NDCG@10 | Recall@20 | NDCG@20 |
| (a) Centric | 0.01127 | 0.01910 | 0.019900 | 0.02685 | 0.06451 | 0.05318 | 0.09319 | 0.06490 |
| (b) Popularity | 0.00819 | 0.01453 | 0.01330 | 0.01906 | 0.03966 | 0.03238 | 0.07058 | 0.04534 |
| (c) MF | 0.00480 | 0.00851 | 0.00872 | 0.01237 | 0.04569 | 0.03700 | 0.07538 | 0.04909 |
| (d) Temp-Fusion | <u>0.01177</u> | <u>0.02005</u> | <u>0.02070</u> | <u>0.02756</u> | **0.06933** | **0.05888** | <u>0.09823</u> | **0.07115** |
| (e) **LLM-TUP** | **0.01320***$^*$ | **0.02174***$^*$ | **0.02225***$^*$ | **0.02932***$^*$ | <u>0.06653</u>$^*$ | <u>0.05470</u>$^*$ | **0.10214***$^*$ | <u>0.06827</u>$^*$ |
| Gain of LLM-TUP vs. (a) | 17% | 14% | 12% | 9% | 3% | 3% | 10% | 5% |

*asterisk indicates that the improvement of the proposed method (LLM-TUP) over the baseline method (Centric) is statistically significant ($p < 0.05$).*

## 4.3 Evaluation Methodology

Given our focus on user profiling in content-based recommendation systems, it is more relevant to evaluate and compare our proposed method against user profiling approaches commonly used in such systems. The centric-based approach, which creates user profiles by averaging item embeddings, serves as a strong baseline for this purpose.

Unlike sequential recommendation models such as [13, 33, 34], which evaluate performance based on next-item prediction tasks, our evaluation methodology focuses on assessing the effectiveness of user profiling for content-based recommendation. In sequential models, the evaluation typically involves auto-regressive prediction of the next interaction within a user's sequence using metrics such as Hit Rate and MRR. In contrast, we evaluate how well the generated overall user profiles rank relevant items among a broader set, using Recall@K and NDCG@K as key metrics. Our evaluation methodology ensures that user representations are optimized for content-based personalization rather than sequence-based transition modeling.

To create a temporally aware experimental setup, each user's interactions were sorted by timestamp, and the data was split into: 1) *Training set*: The earliest 60% of interactions, 2) *Validation set*: The subsequent 20% of interactions, and 3) *Test set*: The most recent 20% of interactions. This time-wise splitting strategy ensures that user preferences evolve naturally from training to test, allowing us to evaluate the models' ability to predict future interactions based on historical behavior.

## 4.4 Experimental Setup

The experiments were implemented using PyTorch [26], and the model was trained with the Adam optimizer. To ensure stable training and prevent overfitting, we utilized early stopping with a patience of 5 epochs, terminating the training if the validation performance did not improve for 5 consecutive epochs. Each model was trained for a maximum of 100 epochs, and the batch size was varied across 512, 1024, and 2048, with the reported results based on a batch size of 2048, which yielded the best performance.

For encoding the natural language representations of user preferences, we used the all-MiniLM-L6-v2 version of SBERT [30], which provides efficient and high-quality semantic embeddings with 384 dimensions. To generate the short-term and long-term natural language profiles of users from their interaction histories, we employed the OpenAI GPT-4o-mini model [1], which excels in generating coherent and contextually rich text representations. The resulting short-term and long-term user embeddings, along with the item embeddings, were all set to have a dimensionality of 384.

The hidden dimension of the multi-layer perceptron (MLP) used for interaction likelihood prediction was set to 128, with a dropout rate of 0.2 applied to mitigate overfitting. During training, we applied negative sampling, generating 5 negative samples for each positive user-item interaction to ensure effective contrastive learning. The learning rate was fixed at 1e-3, which provided stable and reliable convergence.

The experiments were executed on an environment using four NVIDIA A100 GPUs, each with 80GB of memory, enabling efficient parallelization and acceleration of the training process. This setup, along with robust optimization techniques and careful hyperparameter tuning, ensured reliable and reproducible results across all experiments.

### 4.5    Experimental Results & Discussion

This section addresses our first two research questions by assessing the effectiveness of LLM-TUP against multiple baselines across two datasets: Movies&TV and Video Games.

*4.5.1    Answering RQ1: Does LLM-TUP improve recommendation accuracy compared to other baselines?* Table 2 presents the ranking performance of LLM-TUP and baselines on the Movies&TV and Video Games datasets. Across all evaluation metrics, LLM-TUP consistently outperforms Centric, Popularity, and MF baselines, demonstrating its effectiveness in improving ranking accuracy.

On the Movies&TV dataset, LLM-TUP achieves a 17% improvement in Recall@10 and a 14% increase in NDCG@10 over Centric. Similarly, LLM-TUP surpasses Centric by 12% in Recall@20 and 9% in NDCG@20, with all improvements statistically significant ($p < 0.05$).

Among baselines, Temp-Fusion performs the strongest due to its ability to model short-term and long-term preferences. However, LLM-TUP consistently outperforms Temp-Fusion, highlighting the value of LLM-based user profiling in capturing deeper semantic and temporal user behaviors.

On the Video Games dataset, LLM-TUP maintains its best performance on Recall@20. However, Temp-Fusion slightly outperforms LLM-TUP on Recall@10, NDCG@10, and NDCG@20. A detailed analysis of this phenomenon is provided in Section 4.5.3.

*4.5.2    Answering RQ2: How effective is LLM in generating semantically rich and temporally aware user profiles?* To assess the impact of LLM-generated user profiles, we compare LLM-TUP with Temp-Fusion, which represents a traditional content-based method that incorporates temporal modeling but does not leverage LLMs.

On the Movies&TV dataset, LLM-TUP outperforms Temp-Fusion by 12% in Recall@10 and 7% in NDCG@10, demonstrating that LLM-driven user profiles provide more expressive representations than BERT embeddings of item metadata alone. Similar trends are observed in Recall@20 and NDCG@20. These results validate the hypothesis that LLMs enhance both short-term adaptability and long-term preference modeling, making user profiles more semantically meaningful.

For the Video Games dataset, LLM-TUP's performance gains over Temp-Fusion are smaller but still evident in Recall@20, confirming the advantage of LLM-based profiling. However, the narrower performance margins in the Games dataset suggest that the effectiveness of LLM-based profiling depends on dataset characteristics. We analyze these effects in the next section.

*4.5.3    Performance Analysis on the Games Dataset.*  Despite LLM-TUP's overall superiority, its performance on Recall@10, NDCG@10, and NDCG@20 in the Games dataset is comparable to or slightly lower than Temp-Fusion. We identify two key factors contributing to this phenomenon:

Table 3. Recall@K Comparison for Ablation Variants (Movies&TV Dataset)

| Experiment | Recall@10 | Recall@20 |
|---|---|---|
| **Short-Term Only (ST)** | 0.01086 | 0.01932 |
| **Long-Term Only (LT)** | 0.01104 | 0.01946 |
| **General Preferences (No TS)** | 0.01002 | 0.01781 |
| **Dot-Product Scoring (DP)** | 0.00837 | 0.01357 |
| **Proposed (LLM-TUP)** | **0.01320** | **0.02225** |

Table 4. NDCG@K Comparison for Ablation Variants (Movies&TV Dataset)

| Experiment | NDCG@10 | NDCG@20 |
|---|---|---|
| **Short-Term Only (ST)** | 0.01942 | 0.02659 |
| **Long-Term Only (LT)** | 0.01951 | 0.02646 |
| **General Preferences (No TS)** | 0.01829 | 0.02471 |
| **Dot-Product Scoring (DP)** | 0.01277 | 0.01733 |
| **Proposed (LLM-TUP)** | **0.02174** | **0.02932** |

(1) Limited Interaction Histories: As shown in Table 1, users in the Games dataset have significantly fewer interactions (Avg. User Profile Size = 4.55) compared to the Movies dataset (Avg. User Profile Size = 10.28). In sparse scenarios, splitting a short interaction history into short-term and long-term segments may not meaningfully enhance user profiles. Consequently, the benefits of LLM-generated representations and attention-based fusion become less pronounced in this dataset.

(2) Stable User Interests: Unlike movie-watching patterns, which fluctuate based on trends, seasons, or contextual factors, gamers tend to have stable preferences focused on specific genres or franchises. This stability reduces the advantage of dynamically adapting user profiles to short-term shifts, making Temp-Fusion— which directly aggregates temporal interactions— perform competitively.

Despite these nuances, LLM-TUP still achieves the best performance in Recall@20, indicating that its LLM-driven profiling remains beneficial for preference modeling. These findings highlight the importance of dataset characteristics in determining the effectiveness of LLM-based user modeling.

## 5 Ablation Study

To address our third research question (RQ3), we conducted an ablation study to isolate key architectural and design choices, allowing us to assess the contribution of individual components in our proposed method. The experiments include variations that use only short-term or long-term preferences, general preferences without temporal distinction, and an alternative scoring mechanism (dot product) in place of the MLP. Since the results for the Games dataset exhibited a similar pattern, we present only the ablation study results for the Movies dataset due to space constraints. Table 3 and Table 4 illustrate numerical results for Recall@K and NDCG@K. Also, Figure 2 visualizes how these variations affect recommendation quality. To make the analysis more structured and ensure a clear response to RQ3, we provide a separate discussion for each variant below.

### 5.1 Short-Term Preferences Only (ST)

This variant exclusively focuses on short-term user interests. We only use $NL_u^{short}$ as the representation for a user profile. These textual representations are then encoded with BERT to produce user embeddings, which we then combine with item embeddings using an MLP to compute the final scores. We aim to isolate the role of short-term preferences, reflecting rapidly changing or context-specific user behavior, while ignoring broader historical interactions.

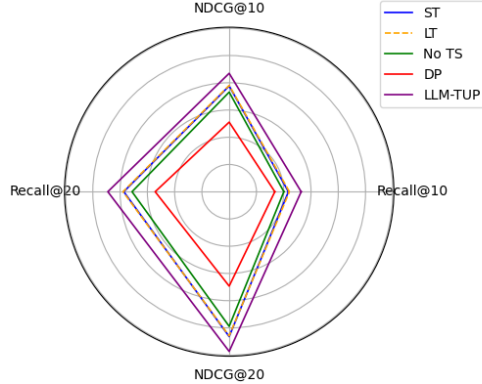

Fig. 2. Comparison of the proposed method against ablated variants on the Movies dataset, illustrating how each approach ranks items (Recall@10, Recall@20) and captures relevance (NDCG@10, NDCG@20) at various cutoff points.

As seen in Tables 3 and 4, ST performs worse than Long-Term Only (LT) suggesting that while short-term interactions contribute to recommendation quality, they alone are insufficient to generate a robust user profile. The performance gap is particularly evident for larger K in Figure 2, where recommended lists must capture more diverse, historically relevant items.

### 5.2 Long-Term Preferences Only (LT)

In this variant, we only use $NL_u^{long}$ as the representation for a user profile which is then transformed into embeddings via BERT and integrated with item embeddings in an MLP. This experiment measures how effectively general, enduring user tendencies can drive recommendations when recent signals are excluded.

In Tables 3 and 4, LT consistently provides respectable scores, emphasizing that many users exhibit relatively stable interests. However, LT lags behind the full model. Figure 2 illustrates this limitation, where LT is consistently outperformed by the full model across all metrics, highlighting the advantage of dynamically integrating both temporal signals rather than relying on long-term behavior alone.

### 5.3 General Preferences without Temporal Distinction (NoTS)

Unlike the ST and LT splits, the "NoTS" setting collapses all user interactions—regardless of recency—into a single textual summary. This removes the temporal component entirely, producing a unified user profile that captures broad preferences but disregards the distinction between recent and historical behaviors. By removing explicit time segmentation, we examine whether temporal separation (i.e., short-term vs. long-term) is essential to robust user modeling.

As seen in Table 3 and 4, NoTS performs better than Dot-Product Scoring (DP) but underperforms compared to both ST and LT, indicating that removing temporal segmentation weakens user profiling effectiveness. Specifically, NoTS achieves a Recall@10 of 0.01002 and an NDCG@10 of 0.01829, which are lower than both ST (0.01086, 0.01942) and LT (0.01104, 0.01951). This suggests that collapsing all interactions into a single representation dilutes critical short-term and long-term signals, leading to less precise personalization. The radar chart in Figure 2 further confirms this, as the NoTS variant fails to match the recall and ranking quality of ST, LT, and LLM-TUP. The decline in performance highlights the importance of explicitly modeling temporal dynamics. Without time-aware segmentation, the model loses the ability to differentiate between persistent user interests and recent contextual shifts, resulting in suboptimal recommendations. This finding reinforces the motivation behind LLM-TUP's attention-based fusion, where the model dynamically balances short-term and long-term preferences per user, yielding stronger ranking accuracy than a uniform profile.

### 5.4 Substitute MLP with Dot Product (DP)

This variant retains the proposed short- and long-term user embeddings but replaces the multi-layer perceptron (MLP) scoring mechanism with a simple dot product between the overall user embedding and the item embedding. We aim to isolate whether the added complexity of a non-linear scoring function (the MLP) significantly contributes to recommendation performance.

Dot product captures only linear relationships between user and item embeddings; hence, nuanced correlations and complex feature interactions are lost. In Tables 3 and 4, DP exhibits lower Recall@K and NDCG@K compared to MLP-based models. Figure 2 underscores this difference, particularly at larger K values, indicating that non-linear transformations yield more accurate item rankings and better personalization.

### 5.5 Proposed Model (LLM-TUP)

Our full approach combines the strengths of short-term and long-term user embeddings, each derived from separate LLM-generated descriptions. These embeddings are fused to form a comprehensive user profile, which is then fed into an MLP alongside item embeddings to produce final recommendation scores. This design captures both immediate user context (via short-term signals) and stable, overarching preferences (via long-term signals). The MLP further models rich, non-linear interactions between user and item representations.

LLM-TUP consistently outperforms all ablation variants in Tables 3 and 4 for both Recall@K and NDCG@K. Its superior performance is visible across the full range of K values in Figure 2, demonstrating how integrating short-term dynamics, long-term stability, and non-linear scoring leads to the most accurate and robust recommendations. Collectively, these results affirm that each aspect—temporal segmentation, dual-profile modeling, and the MLP-based scoring—contributes a distinct and complementary benefit, highlighting the importance of modeling both recent user context and deeper historical trends.

The ablation study addressed RQ3 by systematically confirming that (1) separating short- and long-term preferences, (2) leveraging natural language-based embeddings from LLMs, and (3) employing a non-linear MLP scoring function are all crucial for achieving high-quality recommendations. By combining the adaptiveness of short-term profiles with the stability of long-term profiles and utilizing a non-linear interaction model, LLM-TUP consistently outperforms in Recall@K and NDCG@K across various evaluation scenarios.

## 6 Conclusion

We introduce a novel user modeling approach for content-based recommender systems by incorporating temporal dynamics and leveraging LLMs for expressive user representations. Unlike traditional averaging methods, our approach separates short- and long-term preferences, models them via LLMs, and dynamically fuses them using attention. Experiments show up to 17% improvement in Recall@10 and 14% in NDCG@10 over the centric approach, with statistical tests confirming significance. These results highlight the benefits of semantically rich, time-aware user profiles for personalization.